\documentclass[orivec]{llncs}
\usepackage{makeidx}  
\usepackage{color}
\usepackage{algpseudocode}
\usepackage{listings}

\usepackage[utf8]{inputenc}

\newif\ifdraft\drafttrue
\newif\iffinal\finalfalse
\newif\ifextended\extendedfalse
\newif\ifsubmission\submissiontrue
\newif\ifdotikz\dotikzfalse
\newif\ifinlineref\inlinereffalse

\dotikztrue
\usepackage{graphicx}
\graphicspath{{figures/}}
\usepackage{latexsym}
\usepackage{amsmath}
\usepackage{amstext}
\usepackage{amssymb}
\usepackage{amsfonts}
\usepackage{color}
\usepackage{paralist}
\usepackage{enumerate}
\usepackage{ifpdf}
\usepackage{marvosym}
\usepackage{float}
\usepackage{url}
\usepackage{multirow}
\usepackage{alltt}

\usepackage[lined,linesnumbered,algoruled]{algorithm2e}
\usepackage{algpseudocode}

\usepackage[amsmath]{ntheorem} 

\ifdotikz
\usepackage{pgf}
\usepackage{tikz}
\usetikzlibrary{arrows,positioning,automata,decorations,fit,backgrounds,calc}
\usepgflibrary{shapes.geometric} 
\usetikzlibrary{shapes.geometric} 
\usepgflibrary{decorations.pathmorphing} 
\usetikzlibrary{decorations.pathmorphing} 
\usepgflibrary{decorations.text} 
\usetikzlibrary{decorations.text} 
\usetikzlibrary{matrix}
\pgfrealjobname{main}
\else
\ifpdf
\long\def\beginpgfgraphicnamed#1#2\endpgfgraphicnamed{\includegraphics{#1}}
\else
\usepackage{epsfig}
\long\def\beginpgfgraphicnamed#1#2\endpgfgraphicnamed{\epsfig{file=#1.eps}}
\fi
\fi

\newcommand{\entails}{\Vdash}

\newcommand{\comment}[1]{}

\newcommand{\leanparagraph}[1]{\smallskip\noindent\textbf{#1.} }

{\end{list}}

\newcounter{myenumctr}

\newcommand{\nop}[1]{}

\newcommand{\mi}[1]{\ensuremath{\mathit{#1}}}

\theorembodyfont{\normalfont}
\theoremseparator{.~}
\renewtheorem{example}{Example}

\newcommand{\intensional}{\ensuremath{\cI}}
\newcommand{\extensional}{\ensuremath{\cE}}

\newcommand{\Atoms}{\ensuremath{\cA}}

\newcommand{\Formulas}{\ensuremath{\cF}}

\newcommand{\head}{\ensuremath{\mathit{H}}}
\newcommand{\body}{\ensuremath{\mathit{B}}}

\newcommand{\wfnTime}{\tau}
\newcommand{\wfnTuple}{\#}

\newcommand{\ExtendedAtoms}{\ensuremath{\Atoms^+}}

\newcommand{\opPlaceholder}{\ensuremath{\mathbin{\star}}}

\renewcommand{\vec}[1]{{\mathbf{#1}}}

\newcommand{\backgroundData}{\mathcal{B}}

\mathchardef\minus="2D

\newcommand{\intpr}{v} 

\DeclareMathOperator{\naf}{not}

\newcommand{\window}{\ensuremath{\boxplus}}
\newcommand{\timeWindow}[1]{\ensuremath{\boxplus}^{#1}}
\newcommand{\tupleWindow}[1]{\ensuremath{\boxplus}^{\# #1}}

\SetKwProg{Def}{def}{$\,$:}{}
\SetKwProg{Defn}{defn}{~$=$}{}
\SetKwProg{DefnCustom}{defn}{}{}
\SetKw{Let}{let}
\SetKw{Halt}{halt}
\SetKwInput{KwIn}{Input}
\SetKwFor{ForEach}{foreach}{}{}
\SetKw{Or}{or}
\SetKw{And}{and}
\SetKwIF{If}{ElseIf}{Else}{if}{}{else if}{else}{endif}
\SetKw{Match}{match}
\SetKw{Case}{case}
\SetKw{MapTo}{\ensuremath{\;\;\Rightarrow\;\;}}

\newcommand{\substitution}{\sigma}

\newcommand{\grd}{\mathit{grd}}

\newcommand{\cA}{\ensuremath{\mathcal{A}}}

\newcommand{\cE}{\ensuremath{\mathcal{E}}}
\newcommand{\cF}{\ensuremath{\mathcal{F}}}

\newcommand{\cI}{\ensuremath{\mathcal{I}}}

\newcommand{\bbN}{\ensuremath{\mathbb{N}}}

\newif\ifextended\extendedtrue

\ifextended
\newcommand{\inExtendedVersion}[1]{#1}
\else
\newcommand{\inExtendedVersion}[1]{}
\fi

\newcommand{\squishlist}{
 \begin{list}{$\bullet$}
  { \setlength{\itemsep}{0pt}
     \setlength{\parsep}{3pt}
     \setlength{\topsep}{3pt}
     \setlength{\partopsep}{0pt}
     \setlength{\leftmargin}{1.5em}
     \setlength{\labelwidth}{1em}
     \setlength{\labelsep}{0.5em} } }

     \newcommand{\squishend}{
  \end{list}  }

\begin{document}
\frontmatter          
\pagestyle{headings}  
\addtocmark{Laser} 
\mainmatter              
\title{Expressive Stream Reasoning with Laser\thanks{This work is partially funded by the Dutch public-private research community COMMIT/ and NWO VENI project
639.021.335 and by the Austrian Science Fund (FWF) projects P26471 and W1255-N23.}}
\titlerunning{Stream Reasoning with Laser}  
%
\author{Hamid R. Bazoobandi\inst{1} \and Harald Beck\inst{2}
\and Jacopo Urbani\inst{1}}
\authorrunning{Bazoobandi et al.} 
\tocauthor{Hamid R. Bazoobandi, Harald Beck, Jacopo Urbani} 
\institute{Vrije Universiteit Amsterdam, The Netherlands\\ 
    \email{h.bazoubandi@vu.nl,jacopo@cs.vu.nl},
\and
Institute of Information Systems, Vienna University of Technology\\
  \email{beck@kr.tuwien.ac.at}}

\maketitle              

\begin{abstract} An increasing number of use cases require a timely
  extraction of non-trivial knowledge from semantically annotated data
  streams, especially on the Web and for the Internet of Things
  (IoT). Often, this extraction requires expressive reasoning, which is
  challenging to compute on large streams. We propose Laser, a new
  reasoner that supports a pragmatic, non-trivial fragment of the logic
  LARS which extends Answer Set Programming (ASP) for streams. At its
  core, Laser implements a novel evaluation procedure which annotates
  formulae to avoid the re-computation of duplicates at multiple time
  points. This procedure, combined with a judicious implementation of
  the LARS operators, is responsible for significantly better runtimes
  than the ones of other state-of-the-art systems like C-SPARQL and
  CQELS, or an implementation of LARS which runs on the ASP solver
  Clingo. This enables the application of expressive logic-based
  reasoning to large streams and opens the door to a wider range of
  stream reasoning use cases.
\end{abstract}

\section{Introduction}

The Web and the emerging Internet of Things (IoT) are
highly dynamic environments where streams of data are valuable sources of
knowledge for many use cases, like traffic monitoring, crowd control, security,
or autonomous vehicle control. In this context, reasoning can be applied to
extract 
implicit
knowledge from the stream. For instance, reasoning can be
applied to detect anomalies in the flow of information, and provide clear
explanations that can guide a prompt understanding of the situation.

\leanparagraph{Problem} Reasoning on data streams should be done in a timely
manner~\cite{della2009s,margara_streaming_2014}.
This task is challenging for
several reasons: \emph{First}, expressive reasoning
that supports features for a 
fine-grained control of temporal information
may come
with an
unfavourable computational complexity. This clashes with the requirement of a
reactive system that 
shall work in
a highly dynamic environment. \emph{Second}, the
continuous flow of incoming data 
calls for incremental evaluation techniques that go beyond repeated
querying and re-computation. \emph{Third}, there is 
no consensus on the
formal semantics for the processing of streams 
which
hinders a meaningful and
fair comparison between stream reasoners.

Despite recent substantial progress in the development of stream reasoners, to the best of our
knowledge there is still no
reasoning system
that addresses
all three
challenges. Some systems can handle large streams but do not support expressive
temporal reasoning features~\cite{BarbieriBCVG10,le2011native,AnicicFRS11,sparkwave}. Other
approaches focus on the formal semantics but do not 
provide
implementations~\cite{gebser2013answer}. Finally,
some systems implemented only a particular rule set and cannot be
easily generalized~\cite{hoeksema2011high,urbani2013dynamite}.


%
%
\newcommand{\hackItem}{\noindent $\bullet$\hspace{0.5em}}
\leanparagraph{Contribution} We tackle the above challenges with the following contributions.

\hackItem We present \emph{Laser}, a novel stream reasoning system based
the recent rule-based framework LARS~\cite{bdef15lars}, which extends
Answer Set Programming (ASP) for stream reasoning. Programs are sets of
rules which are constructed on
formulae 
that contain window operators and temporal operators. Thereby, Laser has
a fully declarative semantics amenable for formal comparison.

\hackItem To address the trade-off between expressiveness and data
throughput, we employ a tractable fragment of LARS that ensures
uniqueness of models. Thus, in addition to typical operators and window
functions, Laser also supports operators such as $\Box$, which enforces
the validity over intervals of time points, and $@$, which is useful to
state or retrieve specific time points at which atoms hold.

%

\hackItem We provide a novel evaluation technique which annotates formulae
with two time markers.
%
When a grounding of a formula~$\varphi$ is derived, 
it is annotated with an interval $[c,h]$ from a
\emph{consideration time} $c$ to a \emph{horizon time} $h$, during which
$\varphi$ is guaranteed to hold. By efficiently propagating and removing these
annotations, we obtain an incremental model update that may avoid many
unnecessary re-computations.
%
Also, these annotations enable us to implement a technique similar to the
Semi-Naive Evaluation (SNE) of Datalog programs~\cite{alice} to reduce duplicate
derivations. 

\hackItem We present an empirical comparison of the performance of Laser against
the state-of-the-art engines, i.e., C-SPARQL~\cite{BarbieriBCVG10} and
CQELS~\cite{le2011native} using micro-benchmarks and a more complex program. We
also compare Laser with an open source implementation of LARS which is based on
the ASP solver Clingo to test operators not supported by the other engines.

Our empirical results are encouraging as they show that Laser outperforms the
other systems, especially with large windows where our incremental approach is
beneficial. This allows the application of expressive logic-based reasoning to
large streams and to a wider range of use cases. To the best of our knowledge,
no comparable stream reasoning system that combines similar expressiveness with
efficient computation exists to date.
\ifextended
\else
See~\cite{laser2017arxiv} for an extended version of this paper.
\fi

\section{Theoretical Background: LARS}
\label{sec:background}

As formal foundation, we use the logic-based framework
LARS~\cite{bdef15lars}. We focus on a pragmatic
fragment called \emph{Plain LARS} first mentioned
in~\cite{bde2015-ijcai}. We assume the reader is familiar with basic
notions, in particular those of logic programming.
Throughout, we distinguish \emph{extensional atoms}
$\Atoms^\extensional$ for input and \emph{intensional atoms}
$\Atoms^\intensional$ for derivations.
By $\Atoms = \Atoms^\extensional \cup \Atoms^\intensional$, we denote
the set of~\emph{atoms}.
Basic arithmetic operations and comparisons are assumed to be given in
form of designated extensional predicates, but written with infix
notation as usual. We use upper case letters
$X,Y,Z$
to denote variables, lower case letters $x,y,\ldots$ are for constants, and
$p,a,b,q$ for predicates for atoms.

\begin{definition}[Stream]\label{def:stream} A stream ${S=(T,\intpr)}$ consists
    of a \emph{timeline} $T$, which is a closed interval in $\bbN$, and an
    \emph{evaluation function} ${\intpr : \bbN \mapsto 2^\Atoms}$. The elements
    ${t \in T}$ are called \emph{time points}.  \end{definition}
Intuitively, a stream $S$ associates with each time point a set of atoms.  We
call $S$ a \emph{data stream}, if it contains only extensional atoms.
To cope with the amount of data, one usually considers only recent atoms.  Let
${S=(T,\intpr)}$ and ${S'=(T',\intpr')}$ be two streams s.t. ${S' \subseteq S}$,
i.e., ${T' \subseteq T}$ and ${\intpr'(t') \subseteq \intpr(t')}$ for all ${t'
\in T'}$. Then $S'$ is called a \emph{window} of $S$.
\begin{definition}[Window function]\label{def:window-function} Any (computable)
function $w$ that returns, given a stream $S=(T,\intpr)$ and a time point ${t
\in \bbN}$, a \emph{window} $S'$ of $S$, is called a \emph{window function}.
\end{definition}
In this work, we focus on two prominent sliding windows that select
recent atoms based on time, respectively counting. A sliding time-based window
selects all atoms appearing in the last $n$ time points.
\begin{definition}[Sliding Time-based Window] \label{def:sliding-time-window}
  Let ${S=(T,\intpr)}$ be a stream, ${t \in T=[t_1,t_2]}$ and let
  ${n \in \bbN}$, $n \geq 0$.
  Then
  the \emph{sliding time-based window function $\wfnTime_n$ (for size~$n$)} is
  ${ \wfnTime_n(S,t) = (T',\intpr|_{T'})}$, where ${T'=[t',t]}$ and ${t' =
  \max\{t_1,t-n\}}$.
\end{definition}
Similarly, a sliding tuple-based window selects the last $n$ tuples. We define
the \emph{tuple size $|S|$} of stream $S=(T,\intpr)$ as $|\{ (a,t) \mid t \in T,
a \in \intpr(t)\}|$.
%
%
\begin{definition}[Sliding Tuple-based Window] \label{def:sliding-tuple-window}
  Let ${S=(T,\intpr)}$ be a stream, ${t \in T=[t_1,t_2]}$ and let
  ${n \in \bbN}$, $n \geq 1$. The \emph{sliding tuple-based window
    function $\wfnTuple_n$ (for size~$n$)} is
  \begin{displaymath} \wfnTuple_n(S,t) = \begin{cases} \wfnTime_{t-t'}(S,t) &
  \text{if}~|\wfnTime_{t-t'}(S,t)| \leq n,\\ S' & \text{else,} \end{cases}
  \end{displaymath}
where $t' = \max (\{u \in T \mid |\wfnTime_{t-u}(S,t)| \geq n\} \cup \{t_1\})$
and $S'=([t',t],\intpr')$ has tuple size $|S'|=n$ such that
$\intpr'(u)=\intpr(u)$  for all $u \in [t'+1,t]$ and $\intpr'(t')\subseteq
\intpr(t')$.  \end{definition}
\begin{figure}[t] \centering \footnotesize \beginpgfgraphicnamed{buses-pic0}
    \begin{tikzpicture}[scale=1.20,node distance=0.4cm,>=latex] \draw [->]
        (34.5,0) -- ($(40,0)+(3,0)$) node [anchor=west] {};

      \foreach \i in {35,36,37,38,39,40,41,42} 
      {
        \draw (\i cm,0pt)  node[anchor=north] at (\i,-0.32)
        {$\i$};
        \node at (\i,0) {\scriptsize $\bullet$};
      }
      \foreach \i in {36,38,40}
      {
        \draw (\i cm,0pt)  node[anchor=north] at (\i,-0.32)
        {$\i$};
        \draw [dotted, thick] (\i,0) -- (\i,0.4) ;
      }

      \draw [thick] ($(41,0.3)$) -- ($(41,-0.3)$) ;

      \node (n1) at  ($(36,0.55)$) {$\{a(x_1,y)\}$};
      \node (n2) at  ($(38,0.55)$) {$\{a(x_2,y),b(y,z)\}$};
      \def\closeLabelsOffset{0.0}
      \node (n3) at  ($(40,0.55)-(\closeLabelsOffset,0)$) {$\{a(x_3,y)\}$};

      \def\windowHeight{0.16}
      \draw [thick] (41.0,\windowHeight) -- ($(41.0,\windowHeight)-(3.0,0)$) -- ($(41.0,-\windowHeight)-(3.0,0)$) -- (41.0,-\windowHeight);
    \end{tikzpicture}
  \endpgfgraphicnamed
  \caption{A time (resp. tuple) window of size 3 at $t=41$}
  \label{fig:window}
\end{figure}
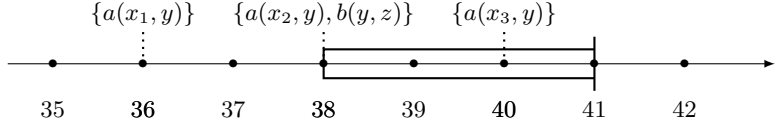
We refer to these windows simply by time windows, resp. tuple windows.
Note that for time windows, we allow size $n=0$, which selects all atoms
at the current time point, while the tuple window must select at least
one atom, hence $n \geq 1$.

Note that we associate with each time point a set of atoms. Thus, for
the tuple-based window, if $[t',t]$ is the smallest timeline in which
$n$ atoms are found, then in general one might have to delete arbitrary
atoms at time point $t'$ such that exactly $n$ remain $[t',t]$.
%
\begin{example}\label{ex:stream}
  Consider a data stream ${D=(T,\intpr_D)}$ as shown in
  Fig.~\ref{fig:window}, where ${T=[35,42]}$ and
  $\intpr_D=\{ 36 \mapsto \{a(x_1,y)\}, 38 \mapsto \{a(x_2,y),b(y,z)\},
  40 \mapsto \{a(x_3,y)\}\}$. The indicated time window of size 3 has
  timeline $[38,41]$ and only contains the last three atoms. Thus,
  the window is also the tuple window of size 3 at 40. Notably,
  $[38,41]$ is also the temporal extent of the tuple window of size 2,
  for which there are two options, dropping either $a(x_2,y)$
  or $b(y,z)$ at time 38.
\end{example}
Although Def.~\ref{def:sliding-tuple-window} introduces nondeterminism,
one may assume a deterministic function based on the implementation at
hand. Here, we assume data is arriving in a strict order from which a
natural deterministic tuple window follows.

\leanparagraph{Window operators $\window^w$}
A window function $w$ can be accessed in rules by window operators.
That is to say, an expression $\window^w \alpha$ has the effect that
$\alpha$ is evaluated on the ``snapshot'' of the data stream delivered
by its associated window function $w$. Within the selected snapshot,
LARS allows to control the temporal semantics with further modalities,
as will be explained below.

%


\subsection{Plain LARS Programs}

Plain LARS programs as in~\cite{bde2015-ijcai} extend normal logic
programs. We restrict here to positive programs, i.e., without negation.

\leanparagraph{Syntax}
We define the set $\ExtendedAtoms$ of \emph{extended atoms} by the grammar
\begin{displaymath}
  a\mid @_t a\mid \window^w @_t a\mid \window^w \Diamond a\mid \window^w \Box a\,,
\end{displaymath}
where ${a\in \Atoms}$ and $t \in \bbN$ is a time point.  The expressions
of form $@_ta$ are called \emph{$@$-atoms}. Furthermore, if
$\opPlaceholder \in \{@_t,\Diamond,\Box\}$, $\opPlaceholder a$ is a
\emph{quantified atom} and ${\window^w \opPlaceholder a}$ a \emph{window
  atom}.
%
We write $\timeWindow{n}$ instead of $\window^{\tau_n}$ for the window
operator using a time window function, and $\tupleWindow{n}$ uses the
tuple window of size $n$.

A \emph{rule} $r$ is of the form~$\alpha \leftarrow \beta_1,\dots,\beta_n$,
where $\head(r)=\alpha$ is the \emph{head} and
$\body(r)=\{\beta_1,\dots,\beta_n\}$ is the \emph{body} of $r$. The head
$\alpha$ is of form $a$ or $@_t a$, where ${a \in \Atoms^\intensional}$,
and each $\beta_i$ is an extended atom.
%
%
A \emph{(positive plain) program} $P$ is a set of rules.
We say an extended atom $\beta$ \emph{occurs} in a program $P$ if ${\beta \in\{ \head(r) \} \cup\body(r)}$ for some
rule ${r\in P}$.
\begin{example}[cont'd]\label{ex:program}
  The rule
  $r = q(X,Y,Z) \leftarrow \timeWindow{3} \Diamond a(X,Y), \tupleWindow{3}
  \Diamond b(Y,Z)$ expresses a query with a join over predicates $a$ and
  $b$ in the standard snapshot semantics: If for some variable
  substitutions for $X,Y,Z$, $a(X,Y)$ holds some time during the last 3
  time points and $b(Y,Z)$ at some time point in the window of the last
  3 tuples, then $q(X,Y,Z)$ is must be inferred.
\end{example}
We identify rules $\alpha \leftarrow \beta_1,\dots,\beta_n$ with implications $\beta_1 \land \dots \land \beta_n \rightarrow \alpha$, thus obtaining by them and their subexpressions the set $\Formulas$ of \emph{formulae}.

\leanparagraph{Semantics}
We first define the semantics of ground programs, i.e., programs without
variables, based on a \emph{structure}
$M=\langle S,W,\backgroundData \rangle$, where $S=(T,\intpr)$ is a
stream, $W$ a set of window functions, and $\backgroundData$ a static
set of atoms called \emph{background data}. Throughout, we use
$W=\{\wfnTime_n, \wfnTuple_n \mid n \in \bbN\}$. We define when extended
atoms $\beta$ (and its subformulae) hold in a structure $M$ at a given
time point $t \in T$ as follows. Let $a \in \Atoms$ and $\varphi$ be a
quantified atom. Then,
\begin{displaymath}
  \begin{array}{l@{\quad\text{iff}\quad}l}
    M,t \entails a  & {a \in \intpr(t)}~\text{or}~{a \in \backgroundData},\\
    M,t \entails \Diamond a & M,t' \entails
                                 a~\text{for some}~t'\! \in T ,\,\\
    M,t \entails \Box a & M,t' \entails a~\text{for all}~t'\! \in T,\\
    M,t \entails @_{t'} a & M,t' \entails a~\text{and}~t' \in T, \\
    M,t \entails \window^w \varphi & M',t \entails \varphi,~\text{where}~M'=\langle w(S,t),W,\backgroundData \rangle.
  \end{array}
\end{displaymath}
For a data stream
${D=(T,v_D)}$, 
any stream ${I=(T,\intpr) \supseteq D}$ that coincides with~$D$ on
$\Atoms^\extensional$ is an \emph{interpretation stream}\/ for $D$, and
a structure ${M=\langle I, W,\backgroundData\rangle}$ an
\emph{interpretation} for $D$.
Satisfaction by $M$ at ${t \in T}$ is as follows: For a rule $r$ of form $\alpha
\leftarrow \beta_1,\dots,\beta_n$, we first define $M,t \models \body(r)$ iff
$M,t \entails \beta_i$ for all $i \in \{1,\dots,n\}$. Then, ${M,t \models r}$
iff ${M,t \entails \alpha}$ or ${M,t \not\models B(r)}$; $M$ is a
\emph{model} of program $P$ (for $D$) at time~$t$, denoted ${M,t \models P}$, if
${M,t \models r}$ for all ${r \in P}$;
and $M$ is \emph{minimal}, if in addition no model ${M'=\langle
S',W,\backgroundData \rangle\neq M}$ of $P$ exists s.t. $S'=(T,\intpr')$
and ${\intpr' \subseteq \intpr}$.
\begin{definition}[Answer Stream] An interpretation stream $I$ is an answer
    stream of program $P$ for the data stream ${D \subseteq I}$ at time $t$, if
    ${M=\langle I,W,\backgroundData \rangle}$ is a minimal model of the
    \emph{reduct} $P^{M,t}=\{r \in P \mid M,t \models \body(r)\}$.
\end{definition}
Note that using tuple windows over intensional data seems neither useful
nor intuitive. For instance, program $P=\{a \leftarrow \tupleWindow{1} \Diamond
b\}$ is inconsistent for a data stream~$D$ at time~$t$, where the last atom
is~$b$, occurring at time~$t-1$: by deriving~$a$ for time~$t$, suddenly~$a$ would
be the last tuple.
%
\begin{proposition} Let $P$ be a positive plain LARS program that employs only
time windows, and tuple window operators only over extensional
atoms. Then,~$P$ always has a unique answer stream.
\end{proposition}

\leanparagraph{Non-ground programs} We obtain the semantics for
non-ground programs in a straightforward way by considering rules with
variables as schematic descriptions of respective ground instantiations.
Substitutions $\sigma$ are defined as usual.
\begin{example}[cont'd]\label{ex:answer-stream} Consider the ground
  program $P$ obtained from rule $r$ of Ex.~\ref{ex:program} by
  replacing variables with constants from the data stream $D$ in
  Ex.~\ref{ex:stream}:
  \begin{displaymath}
    \begin{array}{l@{\quad}r@{~~}c@{~~}l}
    r_1\colon &  q(x_1,y,z) & \leftarrow & \timeWindow{3} \Diamond a(x_1,y), \tupleWindow{3} \Diamond b(y,z)\\
    r_2\colon &  q(x_2,y,z) & \leftarrow & \timeWindow{3} \Diamond a(x_2,y), \tupleWindow{3} \Diamond b(y,z)\\
    r_3\colon &  q(x_3,y,z) & \leftarrow & \timeWindow{3} \Diamond a(x_3,y), \tupleWindow{3} \Diamond b(y,z)\\
    \end{array}
  \end{displaymath}
  At time $t=41$, the time window $\timeWindow{3}$ and the tuple window
  $\tupleWindow{3}$ are identical, as indicated in
  Fig.~\ref{fig:window}, and contain atoms $a(x_2,y)$, $b(y,z)$, and
  $a(x_3,y)$. Consider rule $r_1$. Window atom
  $\timeWindow{3}\Diamond a(x_1,y)$ does no hold, since there is not a
  time point~$t$ in the selected window such that $a(x_1,y)$ holds
  at~$t$. However, the remaining window atoms in $P$ all holds, hence
  the body of rules $r_2$ and $r_3$ hold. Thus, a model of $P$ (for $D$
  at time 41) must include $q(x_2,y,z)$ and $q(x_3,y,z)$. We obtain the
  answer stream
  $D \cup (T,\{ 41 \mapsto \{q(x_2,y,z),q(x_3,y,z)\}\}$).
  %
\end{example}
%
%
\begin{definition}[Output]
  Let $I=(T,\intpr)$ be the answer stream of program $P$ (for a data
  stream $D$) at time $t$. Then, the \emph{output} 
  (of $P$ for $D$ at $t$) is defined by
  $\intpr(t) \cap \Atoms^\intensional$, i.e., the intensional
  atoms that hold at $t$.
\end{definition}
%
%
Given a data stream $D=(T,\intpr)$, where $T=[t_1,t_n]$, we obtain an
\emph{output stream} $S=(T,\intpr)$ by the output at consecutive
outputs, i.e., for each $t' \in T$, $\intpr(t')$ is the output for
$(T',\intpr|_{T'})$, where $T'=[t_1,t']$.
Thus, an output stream is the formal description of the sequence of
temporary valid derivations based on a sequence of answer streams over
a timeline. Our goal is to compute it efficiently.
\begin{example}[cont'd]\label{ex:output}
  Continuing Example~\ref{ex:answer-stream}, the output of $P$ for
  $D$ at $41$ is $\{q(x_2,y,z),q(x_3,y,z)\}$. The output stream $S=(T,\intpr)$ is given by $v = \{ t \mapsto \{ q(x_1,y,z), q(x_2,y,z) \mid t=38,39\} \cup \{ t \mapsto \{q(x_2,y,z), q(x_3,y,z)\} \mid t=40,41,42\}$.
\end{example}
%



\section{Incremental Evaluation of LARS Programs}
\label{sec:evaluation}

In this section, we describe the efficient output stream computation of Laser.
The incremental procedure
consists in
continuously grounding and then annotating formulae with two time points that indicate when and for how long
formulae hold. We thus address two important
sources of inefficiency: grounding (including time
variables) and model computation.

Our work deliberately focuses on exploiting purely sliding
windows. The longer a (potential) step size~\cite{bdef15lars}, the
less incremental reasoning can be applied. In the extreme case of a
tumbling window (i.e., where the window size equals the step size)
there is nothing that can be evaluated incrementally. However, as long
as the two subsequent windows share some data, the incremental
algorithm can be beneficial.
We now give the intuition of our approach in an Example.

%
%
%
\begin{example}[cont'd]\label{ex:algorithm-ideas} Consider again the stream of
    Fig.~\ref{fig:window}, and assume that we are at $t=36$, where $a(x_1,y)$
    appears as first atom in the stream. In rule $r=q(X,Y,Z) \leftarrow
    \timeWindow{3}\Diamond a(X,Y), \tupleWindow{3} \Diamond b(Y,Z)$, the atom
    matches the window atom $\alpha = \timeWindow{3} \Diamond a(X,Y)$, and we
    obtain a substitution $\sigma = \{X \mapsto x_1, Y \mapsto y\}$ under which
    $a(X,Y)$ holds at time $36$. However, for $\alpha$, we can use $\sigma$ for
    the next $3$ time points due to the size of the window and operator
    $\Diamond$. That is, we start \emph{considering} $\sigma$ at time $36$ and
    we have a guarantee that the grounding $\alpha\sigma$ (written postfix)
    holds until time point $39$, which we call the \emph{horizon time}. We thus
    write $\alpha\sigma_{[36,39]}$ for the \emph{annotated ground formula},
    which states that $\timeWindow{3}\Diamond a(x_1,y)$ holds at all evaluation
$t \in [36,39]$, i.e., at $t \in [37,39]$, the neither the grounding nor the
truth of $\timeWindow{3}\Diamond a(x_1,y)$ needs to be re-derived.  \end{example}
\begin{definition}\label{def:annotated-formulae} Let ${\alpha \in \Formulas}$ be
    a formula, and ${c,h \in \bbN}$ such that ${c \leq h}$, and $\sigma$ a
    substitution. Then, ${\alpha \sigma}$ denotes the formula which replaces
    variables in $\alpha$ by constants due to $\sigma$; $\alpha \sigma_{[c,h]}$
is called an \emph{annotated formula}, $c$ is called the \emph{consideration
time} and $h$ the \emph{horizon time}, and the interval $[c,h]$ the \emph{annotation}.
\end{definition}
As illustrated in Ex.~\ref{ex:algorithm-ideas}, the intended meaning of an
annotated formula $\alpha \sigma_{[c,h]}$ is that formula $\alpha \sigma$ holds
throughout the interval $[c,h]$. Annotations might overlap.
\begin{example}\label{ex:overlapping-annotations} Consider an atom $a(y)$
    streams at time points $5$ and $8$. Then, for the formula $\alpha =
    \timeWindow{9} \Diamond a(X)$, we get the substitution $\sigma=\{X \mapsto
    y\}$ and an annotation $a_1=[5,14]$ at $t=5$, and then $a_2=[8,17]$ at
    $t=8$.
  That is to say, $\alpha \sigma = \timeWindow{9} \Diamond a(y)$ holds at all
  time points $[5,14]$ due to annotation $a_1$ and at time points $[8,17]$ due
  to $a_2$, and for each $t \in [8,14]$ it suffices to retrieve one of these
  annotations to conclude that $\alpha \sigma$ holds at $t$.
\end{example}
We note that the tuple window can be processed dually by additionally
introducing a \emph{consideration count} $c_\#$ and a \emph{horizon count}
$h_\#$, i.e., an annotated formula $\alpha\sigma_{[c_\#,h_\#]}$ would indicate
that $\alpha\sigma$ holds when the number of atoms received so far is between
$c_\#$ and $h_\#$. In essence, the following mechanisms work analogously for
time- and tuple-based annotations. We thus limit our presentation to the
time-based case for the sake of simplification.

The consideration time allows us to implement a technique similar to semi-naive
evaluation (SNE) (see, e.g.,\ VLog~\cite{vlog}, RDFox~\cite{rdfox},
Datalog~\cite{alice}) which increases efficiency by preventing duplicate
derivations. Conceptually, SNE is a method which simply imposes that at least
one formula that instantiates the body should be derived during or after the previous
execution of the rule, otherwise the rule would surely derive a duplicate
derivation.
Based on the horizon time, on the other hand, we can quickly determine which
formulae should be maintained in the working memory and which ones can be erased
because they no longer hold. We delete an annotated formula
$\alpha\sigma_{[c,h]}$, as soon as the current time $t$ exceeds $h$.
This way of incrementally adding new groundings and immediately removing
outdated is more efficient than processing all possible groundings.
In
particular, it is more efficient to maintain duplicates with temporal overlaps
as in Ex.~\ref{ex:overlapping-annotations} than looking up existing groundings
and merging their intervals.

\begin{algorithm}[t]
    \footnotesize
    \caption{Evaluation $\mi{Eval}$. INPUT: Data stream $D{=}(T,\intpr_D)$, where $T=[t_1,t_n]$; program $P$. OUTPUT:
    Output stream of $P$ for $D$.}
  \label{alg:overview}

    $S_0, \mi{I}_{0}\leftarrow\emptyset$; (set of ground formulae)\\
    \For{$t_i \in \langle t_1,\ldots,t_n\rangle$}{
        $S_i \leftarrow S_{i-1} \cup \{\, a_{[t_i,t_i]} \mid a \in \intpr_D(t_i)\,\}$;\label{line:init-Si}\\
        $I_i \leftarrow S_i \cup \{\, a_{[c,h]} \mid a_{[c,h]} \in I_{i-1} \land t_i
        \leq h\,\}$;\label{line:init-Ii}\\
        \While{True\label{line:begin-while}}{
            $I \leftarrow I_i$;\label{line:assign-Ii-to-I}\\
            \For{$\alpha \leftarrow \beta_1,\ldots,\beta_n \in P$}{
                \lFor{$j \in \{1,\ldots, n\}$\label{line:body-grd}}{$I_i
                \leftarrow I_i \cup \grd(\beta_j, I, t_1, t_i)$}
                $X \leftarrow \{
                    \alpha\substitution_{[max(c_1,\ldots,c_n),min(h_1,\ldots,h_n)]} \mid \beta\substitution_{[c_1,h_1]},\ldots,\beta\substitution_{[c_n,h_n]} \in
                    I_i$\label{line:line:derivations}\\
                    $\hspace{3em}\land\, c_1,\ldots,c_n \leq t_i \land \bigvee_{j= 1}^n  (c_j {=} t_i)\}$;\\

                $I_i \leftarrow I_i  \cup X \cup \{ \alpha\substitution_{[t_i,t_i]}\mid @_{U}\alpha(\substitution \cup \{U \mapsto t_i\})_{[t_i,h]} \in I_i\}$;\label{line:update-Ii}\\
            }
            \lIf{$I_i = I$\label{line:fixed-point-condition}}{\textbf{break}}
        }\label{line:end-while}
        $\intpr(t_i) = \{ a \in \Atoms^\intensional \mid a_{[c,h]} \in I_i
        \,\land\, c \leq t_i \leq h\};$\label{line:set-output-interpretation} (can be streamed out if needed)
    }
    \Return{$S=(T,v)$}\label{line:return-output-stream}
\end{algorithm}

\leanparagraph{Algorithm 1} We report in Alg.~\ref{alg:overview} the
main reasoning algorithm performed by the system. Sets $I_1,\ldots,I_n$
contain the annotated formulae at times $t_1,\ldots,t_n$; $S_0,I_0$ are
convenience sets necessary for the very first iteration.
At the beginning of each time point $t_i$ we first collect in
line~\ref{line:init-Si} all facts from the input stream. Each atom
$a \in \intpr(t_i)$ is annotated with $[t_i,t_i]$, i.e., its validity
will expire to hold already at the next time point. In
line~\ref{line:init-Ii}, we expire previous conclusions based on horizon
times, i.e., among annotated intensional atoms $a_{[c,h]}$ only those
are retained where $t_i \leq h$. Note that we do not delete atoms from
the data stream.

In lines \ref{line:begin-while}-\ref{line:end-while}, the algorithm
performs a fixed-point computation as usual where all rules are executed until
nothing else can be derived 
(line~\ref{line:fixed-point-condition}).
Lines~\ref{line:body-grd}-\ref{line:update-Ii} describe the
physical execution of the rules and the materialization of the new
derivations. First, line~\ref{line:body-grd} collects all annotated
groundings for extended atoms from the body of the considered rule. We
discuss the details of this underlying function $\grd$ later (see
Alg.~\ref{alg:grd}). In line~\ref{line:line:derivations} we then
consider any substitution for the body that currently holds
$(c_1,\dots,c_n \leq t_i)$. In order to produce a new derivation, we
additionally require at least one formula was not considered in previous
time points ($\bigvee_{j= 1}^n (c_j = t_i)$).

The last condition implements a weak version of SNE, which we call
$\mi{sSNE}$. In fact, it only avoids duplicates between time points and
not within the same time point. In order to capture also this last
source of duplicates, we would need to add an additional internal
counter to track multiple executions of the same rule. We decided not to
implement this to limit the space overhead.

Matching substitutions in line~\ref{line:line:derivations} then are assigned to
the head, where variables which are not used can be dropped as usual.
Notice that consideration/horizon time for the ground head atom is given
by intersection of all consideration/horizon times of the body atoms,
i.e., the guarantee for the derivations is in the longest interval for
which the body is guaranteed to hold.
If the head is of form $@_U \alpha$ and holds now, i.e.,
at~$t_i$, we also add an entry for $\alpha$ to $I_i$
(line~\ref{line:update-Ii}). After the fixed-point computation has terminated
(line~\ref{line:fixed-point-condition}), we can either stream the output at
$t_i$, i.e., $\intpr'(t_i)$ (line~\ref{line:set-output-interpretation}), or
store it for a later output of the answer stream $S$ after processing the entire
timeline (line~\ref{line:return-output-stream}).

\leanparagraph{Algorithm~\ref{alg:grd}} The goal of function $\grd$ is
to annotate and return all ground formulae which hold now or in the
future. Depending on the input formula $\alpha$, the algorithm might
perform some recursive calls to retrieve annotated ground subformulae.
In particular, this function determines the interval $[c,h]$ from a
consideration time $c$ to a horizon time $h$ during which a
grounding holds. It is precisely this annotation which allows us to
perform an incremental computation and avoid the re-calculation of the
entire inference at any time point.

\begin{algorithm}[t]
  \footnotesize
  \caption{Function $\grd$. INPUT: Formula $\alpha$, database $I$,
    beginning time point $t_b$, end time point $t_e$. OUTPUT: Annotated
groundings for $\alpha$.}
  \label{alg:grd}

  \Switch{$\alpha$}{

    \Case{$p(\mathbf{x})\colon$}{
      \Return{$\{ \alpha\substitution_{[c,h]}\mid\alpha\substitution_{[c,h]} \in I\}$\label{line:grd-base-case}}
    }\\

    \Case{$\window^n \beta\colon$}{
        $S \leftarrow \grd(\beta,I, max(0,t_e - n), t_e)$;\\
      \quad \Return{$S \cup \{\alpha\substitution_{[c,\min(c+n,h)]}\mid
        \beta\substitution_{[c,h]} \in S\}$}
    }\\

    \Case{$\Diamond \beta\colon$}{
      $S \leftarrow \grd(\beta,I,t_b,t_e)$; \Return{$S \cup \{\alpha\substitution_{[c,\infty]}\mid
        \beta\substitution_{[c,h]}\in S\}$}
    } \\

    \Case{$\Box \beta\colon$}{
      $S \leftarrow \grd(\beta,I, t_b, t_e)$;\\
      \quad \Return{$S \cup \{ \alpha
        \substitution_{[t_e,t_e]} \mid
        \beta\substitution_{[c_1,h_1]},\ldots,\beta\substitution_{[c_n,h_n]} \in S
        \,\land$\\
        \qquad $c_i \leq c_{i+1},h_{i} \geq c_{i+1} ~\forall 1\leq i < n \,\land $\\
        \qquad$ c_1 \leq
        t_b \land t_e \leq h_n \}\}$}
    }\\
    \Case{$@_{U} \beta\colon$}{
      $S \leftarrow \grd(\beta,I, t_b, t_e)$;\\
      \quad \Return{$S \cup \{ \alpha\sigma_{[c,h]} \mid \alpha\sigma_{[c,h]}
        \in I\}\, \cup$\label{line:at-from-at}\\
      \qquad $\{\alpha\substitution'_{[c,\infty]}\mid\beta\substitution_{[c,h]} \in
        S \land u \in [c,h] \land u \in [t_b,t_e] \land \substitution' = \substitution \cup \{ U \mapsto
        u\}\}$\label{line:at-from-atom}}
    }\\
  }
\end{algorithm}
Function~$\grd$ works by a case distinction on the form of the input
formula $\alpha$, similarly as the entailment relation of the LARS
semantics (Section~\ref{sec:background}).
We explain the first three cases directly based on an example.
\begin{example}[cont'd]\label{ex:grd}
  As in Ex.~\ref{ex:overlapping-annotations}, assume
  $\alpha = \timeWindow{9}\Diamond a(X)$ and input atom $a(y)$ at time
  $5$. Towards annotated groundings of $\alpha$, we first obtain
  the substitution $\substitution=\{X \mapsto y\}$ which can be
  guaranteed only for time point $c=5$ for atom $a(X)$, i.e.,
  $a(X)\substitution_{[5,5]}=a(y)_{[5,5]}$. Based on this, we eventually
  want to compute $\alpha\substitution_{[5,14]}$. This is done in two
  steps. First, the subformula $\beta = \Diamond a(X)$ is agnostic about
  the timeline, and its grounding $\beta\sigma$ gets an annotation
  $[5,\infty]$. The intuition behind setting the horizon time to
  $\infty$ at this point is that $\Diamond \beta$ will always hold as
  soon $\beta$ holds once. The restriction to a specific timeline is
  then carried out when $\beta\sigma_{[5,\infty]}$ is handled in case
  $\timeWindow{9}\beta$, which limits the horizon to
  $\min(c+n,\infty)=14$; any horizon time $h$ received for $\beta$ that
  is smaller than $14$ would remain.
\end{example}
Thus, the conceptual approach of Alg.~\ref{alg:grd} is to obtain the
intervals when a subformula holds and adjust the temporal guarantee
either by extending or restricting the annotation. Since the operator
$\Box$ evaluates intervals, we have to include the boundaries of the
window. That is, if a formula $\timeWindow{n} \Box p(\vec{x})$ must be
grounded, we call $\grd$ in Alg.~\ref{alg:overview} for the entire
timeline $[t_1,t_i]$, where $t_i$ is the current evaluation time. Thus, we
get $t_b=t_1, t_e=t_i$ initially. However, in order for $\Box p(\vec{x})$
to hold under a substitution $\sigma$ within the window of the last $n$
time points, $p(\vec{x}) \sigma $ must hold at every time point
$[t-n,t]$. Thus, the recursive call for $\window^n \beta$ limits the
timeline to $[t_e-n,t_e]$. Then, the case $\Box \beta$ seeks to find a
sequence of ordered, overlapping annotations $[c_1,h_1],\dots,[c_n,h_n]$
that subsumes the considered interval $[t_b,t_e]$. In this case,
$\Box \beta$ holds at $t_e$, but it cannot be guaranteed to hold
longer. Thus, when $\alpha \sigma_{[t_e,t_e]}$ is returned to the case
for $\window^n$, the horizon time will not be extended.
\begin{example}[cont'd]\label{ex:grd-box}
  Consider $\alpha' = \timeWindow{2}\Box a(X)$. Assume that in the
  timeline $[0,7]$ at time points $t=5,6,7$, we received the input
  $a(y)$, hence $\timeWindow{2}\Box a(y)$ has to hold at $t=7$. When we
  call (in Alg.~\ref{alg:overview}) $\grd(\alpha',I,0,7)$, where
  $I=\{ a(y)_{[5,5]}, a(y)_{[6,6]}, a(y)_{[7,7]}\}$, the case for
  $\timeWindow{2}\beta$ will call $\grd(\Box a(X),I,5,7)$. The sequence
  of groundings as listed in $I$ subsumes $[5,7]$, i.e., the scope given
  by $t_b=5$ and $t_e=7$, and thus the case for $\Box$ returns
  $\Box a(y)_{[7,7]}$. The annotation remains for $\alpha'$, i.e.,
  $\grd(\alpha',I,0,7)=\{ \timeWindow{2} \Box a(y)_{[7,7]}\}$.  Note
  when at time $8$ atom $a(y)$ does not hold, neither does
  $\timeWindow{2}\Box a(y)$. Hence, in contrast to $\Diamond$, the
  horizon time is not extended for $\Box$.
\end{example}
With respect to the temporal aspect, the case for~$@$ works similarly as
the one for~$\Diamond$, since both operators amount to existential
quantification within the timeline. In addition $\Diamond$, the
$@$-operator also includes in the time point substitution $U \mapsto u$
where the subformula $\beta$ holds (line~\ref{line:at-from-atom}). In
Line~\ref{line:at-from-at}, we additionally take from $I$ the explicit
derivations for $@$-atoms derived so far.
%

\nop{ ****************
In order to demonstrate (in the form of a sketch of a proof) the
correctness of Algorithm~\ref{alg:overview}, we first characterize the
output of function~$\grd$ (Algorithm~\ref{alg:grd}).
\begin{proposition}\label{prop:grd}
  Given a formula $\alpha$, a database $I$, and time points
  $t_b \leq t_e$, $\grd$ returns all annotated groundings for $\alpha$
  (due to $I$) and its subformulae, reflecting all ground formulae which
  hold now (due to timeline $[t_b,t_e]$) or at time points $t' > t_e$.

  //

  if $\grd(t_1,t_i)$ contains $\alpha_{[c,h]}$, where
  $c \leq t_i \leq h$, then $M,t_i \entails \alpha$
\end{proposition}

[provide high-level idea.]

***********} 

\begin{proposition}\label{prop:term}
  For every data stream $D$ and program $P$, Alg.~\ref{alg:overview} terminates.
\end{proposition}
\inExtendedVersion{
\begin{proof}
  Algorithm~\ref{alg:overview} contains four loops. The for-loop
  starting in line 2 ranges over finitely many time points. To see that
  the inner while-loop (starting at line 5) always terminates, we argue
  that $I_i=I$ eventually holds: Initially, the identity is given in
  line 6. Next, a finite (and fixed) number of rules is iterated in line
  7. For the considered rule head, new groundings $X$ will be derived
  based on $I_i$ and then added to $I_i$. For every iteration in the
  while-loop, each rule is considered only once, and there each body is
  considered only once in the for-loop in line 8. It remains to argue
  that $I_i$ cannot grow indefinitely, i.e., that in some iteration, no
  new groundings for body elements are derivable anymore (and thus no
  new rule firings apply). Leaving aside annotations, the case is as
  usual, i.e., the condition follows from the fact that the set of input
  atoms is finite, and thus also the set of possible substitutions;
  assuming usual safety restriction that there is no recursion through
  arithmetic expressions or that the set of possible terms (including
  numbers) is finite. (A practical assumption is, e.g., to limit numbers
  to those in the timeline.) Thus, possible time references are also
  bounded, and thus there is a finite number of annotations that can be
  assigned to any substitution. Consequently, at some point, no further
  combination of a substitution $\sigma$ and annotation $[c,h]$ can be
  assigned to a formula, hence the condition in $I_i=I$ holds and the
  while-loop terminates.
  %
  \hfill$\Box$
\end{proof}
} 

\begin{theorem} Let $P$ be a positive plain LARS program, $D$ be a data
  stream with timeline $T=[t_1,t_n]$. Then, $S$ is the output stream of
  $P$ for $D$ iff $S=\mi{Eval}(D,P)$.
\end{theorem}

\inExtendedVersion{
\begin{proof}[Sketch]
  To show that $\mi{Eval}(D,P)$ corresponds to the output stream of $P$
  for $D$, we first observe that $\mi{Eval}$ always terminates
  (Proposition~\ref{prop:term}).
  Towards the correctness, we recall that the output stream (for
  timeline $T=[t_1,t_n]$) is defined as the sequence of consecutive
  outputs for time points $t_1,\dots,t_n$. This is accounted for by the
  for-loop from lines 2-16, where line 15 assembles the current
  output. We obtain the output stream as follows.

  In the first iteration we have $t_i=t_1$. In line 3, $S_i=S_1$ is
  initialized with the atoms appearing in the data stream at time $t_i$
  ($S_0$ is empty); $S_i$ is copied to $I_i$, since all atoms have the
  annotation $[t_1,t_1]$. The while-loop starting in line 5 then
  accounts for the fixed point computation for rule derivations, i.e.,
  we will add to $I_i$ atoms and @-atoms (more precisely, substitutions
  with annotations that reflect ground atoms and ground @-atoms)
  corresponding to heads of rules which hold. Since we consider positive
  programs, the order in which rules are considered is arbitrary, hence
  we start a for-loop in line 7 for rule traversal.

  In line 8, we collect substitutions for body formulae (and their
  subformulae) due to the current database $I$ (data and derivations so
  far) by means of $\grd$ (Algorithm~\ref{alg:grd}) which we discuss
  below. By definition, an element
  $\beta\sigma_{[c,h]} \in \grd(\beta,I,t_1,t_i)$ means that formula
  $\beta$ (which may be ground or non-ground), grounded by substitution
  $\sigma$, is guaranteed to hold from time $c$ to time $h$ due to the
  current database $I$, given a timeline from $t_1$ to $t_i$.

  Based on the derived evidence for body elements, we check in line 9
  whether the considered rule $\alpha \leftarrow \beta_1,\dots,\beta_n$
  fires, i.e., whether there is a substitution $\sigma$ for
  $\beta_1,\dots,\beta_n$ with annotations $[c_1,h_1],\dots,[c_n,h_n]$,
  respectively, that can already be considered
  ($c_1,\dots,c_n \leq t_i$).  In this case, the head $\alpha$ can be
  derived; more precisely, the substitution carries over for the rule
  head with an annotation obtained as intersection of the body
  annotations, i.e., the latest consideration time and the earliest
  horizon time. This is largest interval for which the body is
  guaranteed to hold. To additional condition
  $\bigvee_{j= 1}^n (c_j = t_i)$ ensures that at least one of the
  consideration times $c_1,\dots,c_n$ matches the current time point,
  i.e., this is only an optimization step to avoid reconsideration of
  existing derivations at later time points. Accordingly, annotated
  formulae are stored in $X$.

  Next, line 11, these collected inferences are added to $I_i$ along
  with derivations for @-atoms which start to hold now. Note that the
  database entry $@_{t_i}\alpha_{[t_i,h]}$ means that from time point
  $t_i$ to time point $h$, $\alpha$ holds \emph{for} time point $t_i$,
  i.e., the mapping $t_i \mapsto \alpha$ is guaranteed to hold within
  $[t_i,h]$. Formally, ${M,t' \entails @_{t_i}\alpha}$ (in the according
  structure $M$), where ${t' \in [t_i,h]}$. However, at any time point
  ${t' \in [t_i+1,h]}$ we do not obtain a guarantee for $\alpha$ to hold
  \emph{at} $t'$ (from this entry); i.e., ${M,t' \entails \alpha}$ for
  ${t' \in [t_i+1,h]}$ is not implied. We observe that
  $@_{t_i} \alpha_{[t_i,h]}$ only implies ${M,t_i \entails \alpha}$;
  thus $\alpha\sigma$ is annotated with $[t_i,t_i]$. This concludes the
  while loop and we end up with annotated formulae
  $\alpha\sigma_{[c,h]}$ in $I_i$ for currently derivable information
  based on the fixed point computation.
  Finally, we get by all currently derivable atoms the output $v(t_i)$
  at $t_i$, i.e., annotated atoms of form $a_{[c,h]}$ in $I_i$, where
  $a$ is intensional and the current time $t_i$ is within $[c,h]$.

  For the remaining iterations of the outer for-loop
  ($t_i=t_2,\dots,t_n$), observe that Line 3 simply expands the history
  of the input stream, and line 4 keeps only those derivations which
  have not yet expired, i.e., conclusions from previous iterations that
  do not need to be recomputed.

  It remains to show that $\grd$ computes the correct substitutions and
  formulae due to the LARS semantics. Let $\alpha$ be a formula, $I$ be
  the current database and the timeline be $[t_b,t_e]$ (which is always
  $[t_1,t_i]$). We want to obtain annotated substitutions for $\alpha$
  and its subformulae that reflect when an according ground formula is
  guaranteed to hold. We do so by a case distinction:

  For the case of a predicate $p(\vec{x})$, we simply return the
  elements of the database obtained so far. In case of a window atom
  $\window^n \beta$, we first retrieve recursively the groundings in a
  timeline narrowed down as stated by the window, i.e., for the interval
  $[\max(0,t_e-n),t_e]$. The $\max$ serves to prevent the extreme case
  where the window would reach back beyond the end of the
  stream. Substitutions $\sigma$ returned for $\beta$ are applicable for
  the window formula, which narrows down the temporal scope of the
  annotation to $[c,\min(c+n,h)]$, i.e., the consideration time remains
  and the horizon time is either carried over (if it is natural number)
  or determined at this point due to the window length (if infinite; see
  cases below).

  For case $\Diamond \beta$, we likewise determine the groundings for
  $\beta$ in the entire considered (global) interval $[t_b,t_e]$. As
  such, for any grounding $\beta\sigma_{[c,h]}$ retrievable for $\beta$,
  $\alpha=\Diamond\beta$ would hold forever, i.e., never expire. Thus,
  we assign $[c,\infty]$, and only the window operator as discussed
  above will then limit the horizon time due to its length. (Note that a
  formula of form $\Diamond \beta$ occurs only in the scope of a window
  operator.)

  The case $\Box \beta$ is dual, i.e., a grounding $\beta\sigma$ has to
  hold at all time points in a considered interval, and no guarantee is
  obtained for the next time point. This is reflected in lines 7-9 in
  Algorithm~\ref{alg:grd}: We look for a substitution $\sigma$ for
  $\beta$ with consecutive ($c_i \leq c_{i+1}$) and overlapping
  ($h_i \geq c_{i+1}$) annotations $[c_1,h_1],\dots,[c_n,h_n]$ such that
  the considered interval $[t_b,t_e]$ is included in their union. Then,
  we infer that ${\alpha=\Box \beta}$ holds now (${t_e = t_i}$ of
  Algorithm~\ref{alg:overview}), but no further guarantee is available,
  hence the annotation $[t_e,t_e]$.

  For case $@_U\beta$ we again first retrieve groundings
  $\beta\sigma_{[c,h]}$ for subformula $\beta$. Then, for all time
  points $u \in [c,h]$, due to the definition of the @-operator,
  $@_u\alpha$ holds for every $u$ is contained in the considered
  timeline $[t_b,t_e]$. That is, we get for any such $u$ a new
  substitution $\sigma'$ by adding to $\sigma$ the binding
  $U \mapsto u$. Note that $@$ is, like $\Diamond$, an existential
  quantification over the timeline, which additionally stores the time
  point. Thus, similarly as in the case for $\Diamond \beta$,
  $@_U\beta\sigma'$ gets the annotation $[c,\infty]$ and the outer
  window operator will limit the scope of the horizon based on its
  length.

  This concludes the proof sketch.
  \hfill$\Box$
\end{proof}
} 

\leanparagraph{Tuple-based windows} As noted earlier, our
annotation-based approach based on consideration time $c$ and horizon
time $h$ works analogously from the tuple-based window by additionally
working with a consideration count $c_{\#}$ and a horizon count $h_\#$ for
every ground formula. Each formula can then hold and expire in only one
of these dimensions, or both of them at the same time.
\begin{example}
  Consider again rule $r$ from Ex.~\ref{ex:algorithm-ideas}. When
  $b(y,z)$ streams in at time $38$ as third atom, we obtain an annotated
  ground formula $\tupleWindow{3} \Diamond b(y,z)_{[3\#,5\#]}$. That is,
  when the fourth and fifth atoms stream in, regardless at which time
  points, $\tupleWindow{3} \Diamond b(y,z)$ is still guaranteed to hold.
\end{example}
\leanparagraph{Adding negation} Notably, our approach can be extended
for handling negation as well. In plain LARS as defined
in~\cite{bde2015-ijcai}, extended atoms $\beta$ from rule bodies may
occur under negation. We can, however, instead assume negation to occur
directly in front of atoms: Due to the FLP-semantics~\cite{FLP04} of
LARS~\cite{bdef15lars}, where ``$\naf$'' can be identified with $\neg$,
we get the following equivalences for both
$w \in \{\wfnTime_n, \wfnTuple_n \}$:
${\neg \window^w \Diamond a(\vec{x})} \equiv {\window^w \Box \neg
  a(\vec{x})}$ and
${\neg \window^w \Box a(\vec{x})} \equiv {\window^w \Diamond \neg
  a(\vec{x})}$. The case is more subtle for $@$, since
$@_t \neg a(\vec{x})$ implies that $a(\vec{x})$ is false. However, due
to the definition of $@$, $\neg @_t a(\vec{x})$ can also hold if $t$ is
not contained in the considered timeline. Thus, the equivalence
${\neg \window^w @_t a(\vec{x})} \equiv {\window^w @_t \neg a(\vec{x})}$
(necessarily) holds only if the timeline contains~$t$.
%
This assumption is safe when we assume that the timeline always covers
all considered time points.

Our approach extends naturally to a variant of plain LARS where negation appears
only in front of atoms: In addition to the base case $p(\vec{x})$ in
Line~\ref{line:grd-base-case} in Alg.~\ref{alg:grd} we must add a case for a
negative literal $\ell=\neg p(\vec{x})$. Using standard conventions, we then
have to consider all possible substitutions $\substitution$ for variables
in~$\vec{x}$ that occur positively in the same rule $r$, such that
$p(\vec{x})\substitution$ does not hold.

We obtain a fragment that is significantly more expressive, but results
in having multiple answer streams in general: note that plain LARS
essentially subsumes normal logic programs, and the program
$a \leftarrow \naf b;~ b \leftarrow \naf a$ has two answer sets $\{a\}$
and $\{b\}$. Analogously, we get multiple answer streams by allowing
such loops through negation. To retain both unique model semantics and
tractability, we propose restricting to stratified negation, i.e.,
allowing negation but no loops through negation. Then, we can add to
Alg.~\ref{alg:overview} an additional for-loop around
lines~\ref{line:assign-Ii-to-I}-\ref{line:fixed-point-condition} to
compute the answer stream stratum by stratum bottom up as usual.
In fact, our implementation makes use of this extension.


\section{Evaluation}

We evaluate the performance of
\textit{Laser}\footnote{\url{https://github.com/karmaresearch/Laser}} on
two dimensions: First, we measure the impact of our incremental procedures on
several operators by micro-benchmarking the system on special single-rule
programs. Second, we compare the performance against the state of the art on
more realistic programs.

\leanparagraph{Streams} Unfortunately, we could not use some well-known stream
reasoning benchmarks (e.g., \emph{SRBench}~\cite{zhang2012srbench},
\emph{CSRBench}~\cite{dell2013correctness} \emph{LSBench}~\cite{le2012linked},
and \emph{CityBench}~\cite{ali2015citybench}) because
\begin{inparaenum}[(i)]
\item we need to manually
change the window sizes and the speed of stream in order to benchmark our incremental approach, but this is not often supported in these benchmarks;
\item in order to be
effective, a micro-benchmark needs to introduce as little overhead as possible;
\item we needed to make sure that all reasoners return the same results for a fair
comparison, and this was easier with a custom data generator that we
wrote for this purpose.  
\end{inparaenum}


\leanparagraph{State-of-the-art} In line with current literature, we
selected \emph{C-SPARQL}~\cite{BarbieriBCVG10}, and
\emph{CQELS}~\cite{le2011native} as our main competitors. 
  For LARS operators that are not supported by these engines, we compare
  Laser with \emph{Ticker}~\cite{ticker-tplp17}, another recent engine
  for (non-stratified) plain LARS
  programs.\footnote{\url{https://github.com/hbeck/ticker}} Ticker comes
  with two reasoning modes, a fully incremental one, and another one
  that uses an ASP encoding which is then evaluated by the ASP solver
  Clingo~\cite{GebserKKS14}. The incremental reasoning mode was not
  available at the time of this evaluation. Thus, our evaluation against
  Ticker concerns only the reasoning mode which is based on Clingo.
%

\leanparagraph{Data generation} Unfortunately, each engine has its own routines
for reading the input. As a result, we were compelled to develop custom data
generators to guarantee fairness. A key problem is that CQELS
processes every new data item immediately after the arrival in contrast to Laser
and C-SPARQL that process them in batches. Hence, to control the
number of triples that stream into CQELS, and make sure that all engines receive
equal number of triples at every time point, we configured each data generator
to issue a triple at calculated intervals.  For this same reason, we report the
evaluation results as the average runtime per input triple and not runtime per
time point.

\leanparagraph{Experimental platform} The experiments were performed on a
machine with 32-core Intel(R) Xeon(R) 2.60GHz and 256G of memory. We used Java
1.8 for C-SPARQL and CQELS and PyPy 5.8 for Laser.  We set the initial Java heap
size to 20G and increase the maximum heap size to 80G to minimize potential negative effects of JVM garbage collection.
For Ticker we used Clingo 5.1.0.



\leanparagraph{Window-Diamond} The standard snapshot semantics employed in C-SPARQL and CQELS selects recent data and then abstracts away the timestamps. In LARS, this amounts to using $\Diamond$ to existentially quantify within a window.
Here, we evaluate how efficiently each engine can evaluate 
this case.

We use the rule
$q(A,B) \leftarrow \window^n\Diamond p(A,B)$, where a predicate of form $r(A,B)$ corresponds
to a triple $\langle A, r, B \rangle$. The window size and the \emph{stream rate} (i.e. the number of atoms streaming in the system at every time point) are
the experiment parameters. We create a number of
artificial streams which produces a series of unique atoms with
predicate $p$ at different rates; we vary window
sizes from 1sec to 80secs and the stream rate from 200 to 800 triples
per second (t/s).

\begin{figure}[tb]
    \centering
    \includegraphics[width=1.0\textwidth]{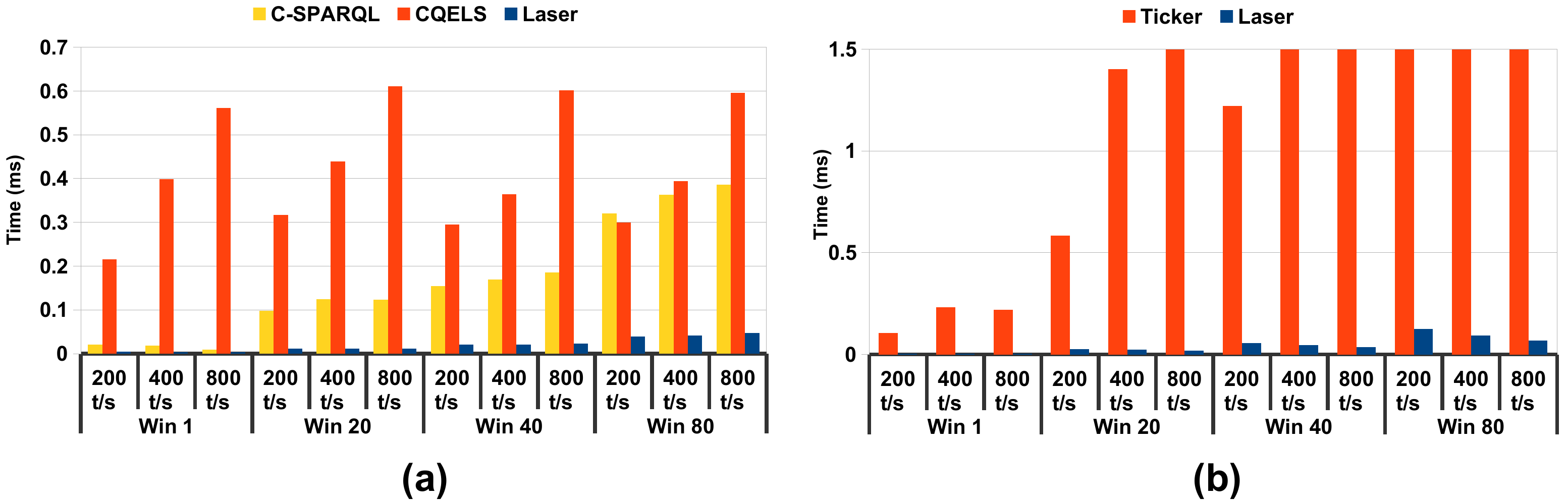}
    \caption{\textbf{(a)} Avg runtime of $\Diamond$
    \textbf{(b)} and of $\Box$ on multiple window sizes and stream rates.}
    \label{fig:diamond_box}
\end{figure}

Fig.~\ref{fig:diamond_box}(a) reports the average runtime per input triple for
each engine.  The figure shows that Laser is faster than the other engines.
Furthermore, we observe that average runtime of Laser grows significantly slower with the window size
 as well as with the stream rate. Here, incremental reasoning clearly is beneficial.

\leanparagraph{Window-Box} The Box operator is not available in C-SPARQL
and CQELS. The semantics of $\Box$ (as well as $@$) may be encoded using explicit timestamps in additional triples but the languages themselves do not directly support it. Therefore, we evaluate the performance of Laser against
Ticker. Similar to the experiments with $\window\Diamond$, we employ the rule
$q(A,B) \leftarrow \window^n\Box p(A,B)$. The experimental settings are
similar to the previous experiment and results are reported in
Fig.~\ref{fig:diamond_box}(b), showing that Laser was
orders of magnitude faster than Ticker.
Notice that with $\Box$ we cannot extend the horizon time, therefore the
incremental evaluation cannot be exploited.
Thus, the performance gain stems from maintaining existing substitutions instead of full recomputations.

\leanparagraph{Data joins} We now focus on a rule which requires a data join.
The computation evaluates the rule $q(A,C)\leftarrow
\window^n\Diamond p(A,B),\window^n\Diamond p(B,C)$ with different window
sizes/stream rates. This program adds the crucial operation of performing a
join.
 From the results reported in
Fig.~\ref{fig:join_rules}(a), we observe the following:

\begin{figure}[tb] \centering
\includegraphics[width=1.0\textwidth]{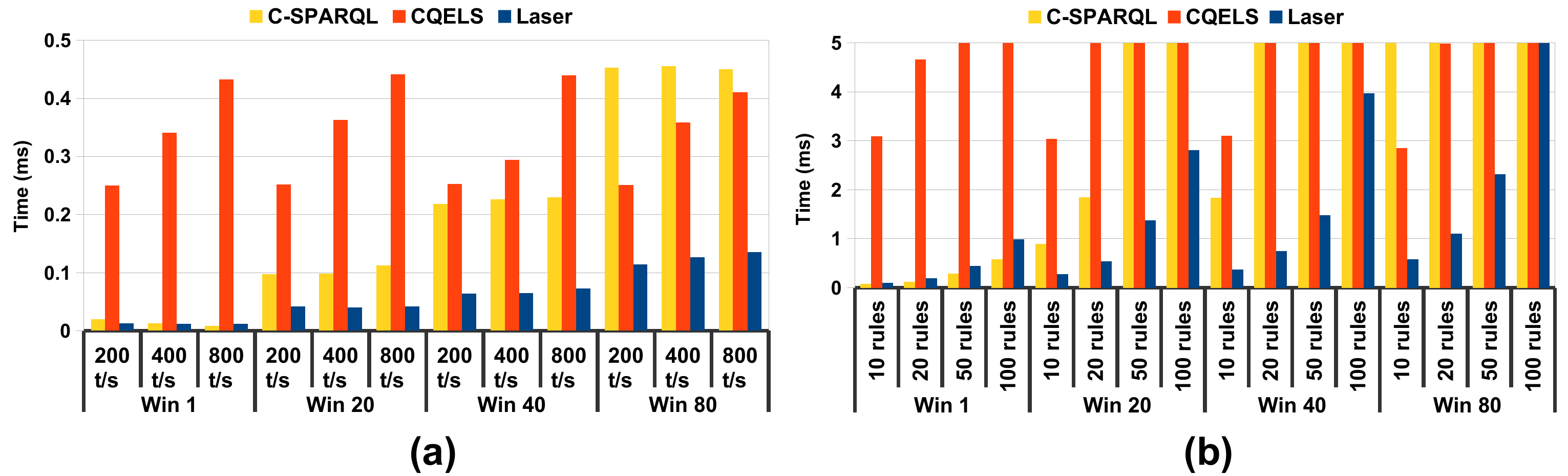} \caption{\textbf{(a)}
Avg. runtimes when the rule requires a data join \textbf{(b)} Avg. runtimes with
multiple rules.} \label{fig:join_rules} \end{figure}

\begin{inparaenum}[(i)] \item Laser is significantly faster that CQELS and C-SPARQL
  with all configurations of window and stream sizes.
\item The difference becomes bigger for larger window sizes for which the benefit of incremental evaluation increases.
\end{inparaenum}

We profiled the execution of Laser with the larger windows and stream sizes and
discovered that only about half of the time is spent on the join while half is
needed to return the results. We also performed an experiment where we
deactivated sSNE and did a normal join instead. We observed that sSNE is slightly
slower than the normal join with small window sizes, but as the size of windows
and stream rate increase, sSNE is significantly faster. In the best
case, the activation of sSNE produced a runtime which was 10 times lower.

\leanparagraph{Evaluating multiple rules} We now evaluate the performance of
Laser in a situation where the program contains multiple rules. In C-SPARQL or
CQELS, this translates to a scenario where there are multiple standing queries.
To do so, we run a series of experiments where we changed the number of rules
and the window sizes (stream rate was constant at 200 t/s). To that end, we
utilize the same rule that we used in the \emph{data join} benchmark with the same data
generator. Fig.~\ref{fig:join_rules}(b) presents the average runtime (per
triple). We see that also in this case Laser outperforms both C-SPARQL and
CQELS, except in the very last case where all systems did not finish on time.

\leanparagraph{Cooling use case} So far we have evaluated the performance using
analytic benchmarks. Now, we measure the performance of Laser with a
program
that deals with a cooling system.
The program of Fig.~\ref{fig:lars_program_rules} determines based on a
water temperature stream whether the system is working under normal
conditions, or it is too hot and produces steam, or is too cold and the
water is freezing.
  
The system also reports temperature readings that are either too high or
too low.
Note that both $@$ (especially in the rule head) and $\Box$ go beyond
standard stream reasoning features. It is not possible to directly translate
this program into C-SPARQL or CQELS queries, so we can only compare the
performance of Laser with Ticker. In this case, the data generator
produces a sequence of random temperature readings. Like before, we
gradually increased the window size and stream rate. The results, shown in
Fig.~\ref{fig:lars_program_performance}, indicate that Laser is
considerably faster than Ticker and can maintain a good response time
($\leq 100\mu$sec) even when the readings come with high frequency (800
t/s).

\begin{figure*}[t]
    \centering
    \scriptsize
    \begin{displaymath}
        \begin{aligned}
            & r_1:\; @_T \;steam(V)  \leftarrow  \;\! \window^{n} @_T \;\! temp(V),~ {V \geq 100} &&
            r_6:\; normal \leftarrow \window^{n}\Box isLiquid \\
            & r_2:\; @_T \; liquid(V)  \leftarrow  \;\! \window^{n} @_T \;\! temp(V),~ {V \geq 1},~ {V < 100} &&
            r_7:\; freeze \leftarrow \naf alarm \;\!, \naf normal \\
            & r_3:\; @_T \;\! isSteam \leftarrow  \window^{n} \;\! @_T\;\! steam(V) &&
            r_8:\; veryHot(T) \leftarrow \window^{n} \;\! @_T \;\! steam(V), {V \geq 150}  \\
            & r_4:\; @_T \;\! isLiquid \leftarrow  \window^{n} \;\! @_T\;\! liquid(V) &&
            r_9:\; veryCold(T) \leftarrow  \window^{n} \;\! @_T\;\! liquid(V),{V = 1} \\
            & r_5:\; alarm \leftarrow \window^{n} \Box \;\! isSteam &&
        \end{aligned}
    \end{displaymath}
    \vspace{-2ex}
    \caption{Program for a cooling system monitoring.}
    \label{fig:lars_program_rules}
\end{figure*}

\begin{figure}[tb]
    \centering
    \includegraphics[width=0.5\textwidth]{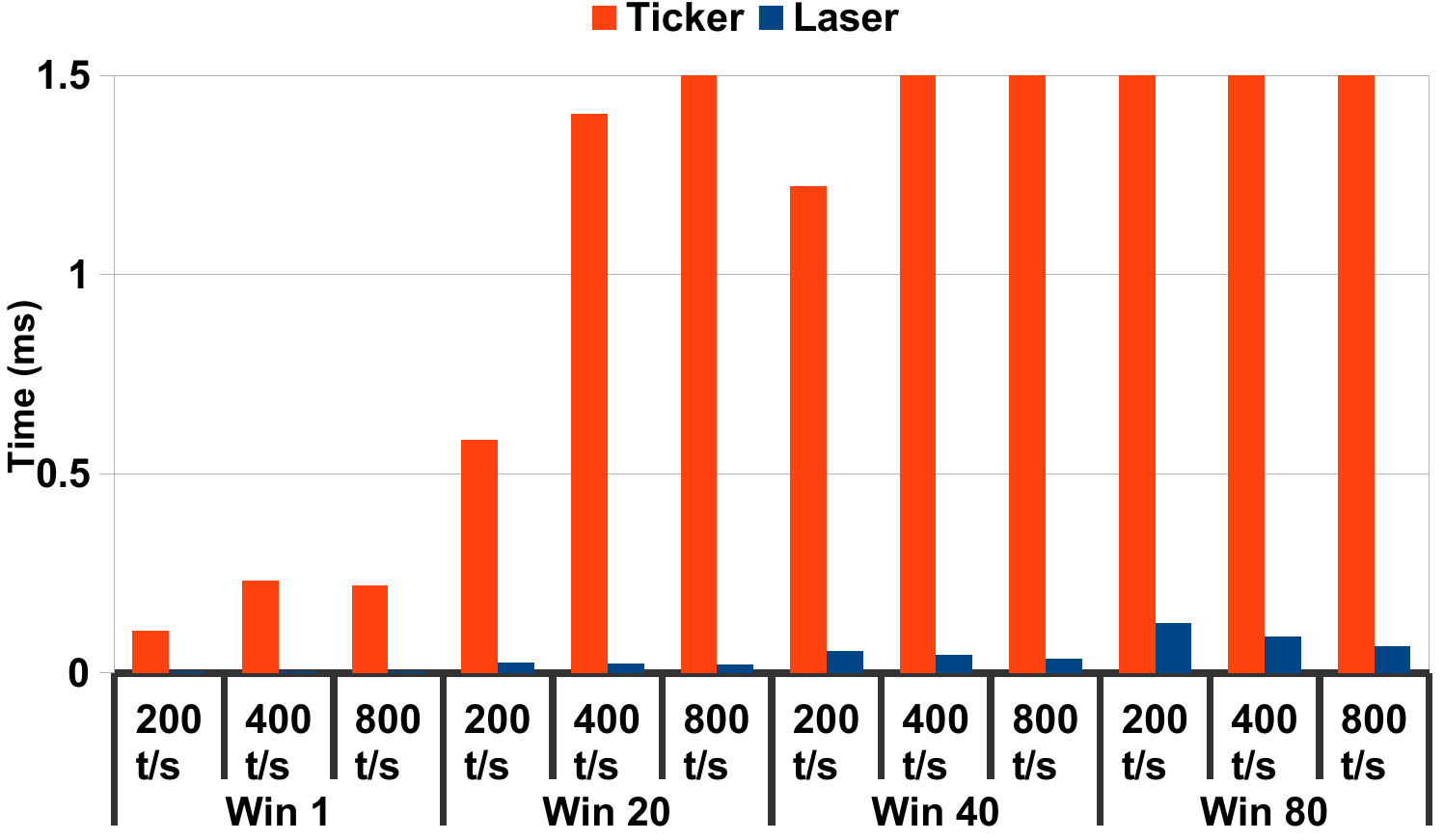}
    \caption{Average execution time per atom of Lars program in
    Fig~\ref{fig:lars_program_rules}.}
    \label{fig:lars_program_performance}
\end{figure}

\section{Related Work and Conclusion}
\label{sec:related}

\leanparagraph{Related Work} The vision of stream reasoning was proposed by Della
Valle et al. in~\cite{della2009s}. Since then, numerous publications have
studied different aspects of stream reasoning such as: extending SPARQL for
stream querying~\cite{barbieri2009c,le2011native}, building stream
reasoners~\cite{barbieri2009c,le2011native,mileo2013streamrule}, scalable stream
reasoning~\cite{hoeksema2011high}, and ASP models for stream reasoning~\cite{gebser2013answer}.
However, due to lack of standardized
formalism for RDF stream processing, each of these engines provide a different
set of features, and results are hard to compare. A survey of these techniques
is available at~\cite{margara_streaming_2014}. Our work differs in the sense
that it is based on LARS~\cite{bdef15lars}, one of the first formal semantics
for stream reasoning with window operators.

An area closely related to stream processing is incremental reasoning,
which has been the subject of a large volume of
research~\cite{motik2015incremental,urbani2013dynamite}. In this
context,~\cite{barbieri2010incremental} describes a technique to add
expiration time to RDF triples to drop them when the are no longer valid. Nonetheless, this approach does not support
expressive operations such as $\Box$ and $@$ that our engine supports. In a
similar way,~\cite{le2017operator} proposes another incremental
algorithm for processing streams which again boils down to efficiently
identifying expired information. We 
showed that our approach outperforms their
work. Next,~\cite{bde2015-ijcai} 
proposes a 
technique to incrementally update an answer stream of a so-called
s-stratified plain LARS program by extending truth maintenance
techniques. While~\cite{bde2015-ijcai} focuses on multiple models, we aim at
highly efficient reasoning for use cases that guarantee single models.
Similarly, the incremental reasoning mode of
  Ticker~\cite{ticker-tplp17} focuses on model maintenance but not on
  high performance.
Stream reasoning based on ASP was also explored in a probabilistic context~\cite{NicklesM15} which however did not employ windows.

\leanparagraph{Conclusion} We presented Laser, a new stream reasoner that is
built on the rule-based framework LARS. Laser distinguishes itself by
supporting expressive reasoning without giving up efficient computation. Our
implementation, freely available, has competitive performance with the current
state-of-the-art. This indicates that expressive reasoning is possible also on
highly dynamic streams of data. Future work can be done on several fronts:
Practically, our techniques extend naturally to further windows
operators such as tumbling windows or tuple-based windows with
pre-filtering. From a theoretical perspective, the question arises which
variations or more involved syntactic fragments of LARS may be
considered that are compatible with the presented annotation-based
incremental evaluation. Moreover, our support of stratified negation is
prototypical and can be made more efficient. More generally,
investigations on the system-related research question of reducing the
runtimes even further are important to tackle the increasing number and
volumes of streams that are emerging from the Web.


\ifinlineref

\else
\bibliographystyle{plain}
\bibliography{references}
\fi

\end{document}